\documentclass[preprint,12pt,authoryear]{elsarticle}

\usepackage{amssymb}
\usepackage{amsmath}
\usepackage{booktabs} 
\usepackage{hyperref}
\usepackage{adjustbox}
\usepackage{lineno}

\journal{Atmospheric Environment}
\begin{document}

\begin{frontmatter}



\title{Dispersion in an array of buildings in stable and convective atmospheric conditions}


\author[author]{Davide Marucci}
\author[author]{Matteo Carpentieri\corref{mycorrespondingauthor}}
\ead{m.carpentieri@surrey.ac.uk}
\address[author]{EnFlo, Department of Mechanical Engineering Sciences, University of Surrey, Guildford, Surrey GU2 7XH, UK}
\cortext[mycorrespondingauthor]{Corresponding author}

\begin{abstract}
	Wind tunnel experiments were conducted to study the impact of atmospheric stratification on flow and dispersion within and over a regular array of rectangular buildings. Three stable and two convective incoming boundary layers were tested with a Richardson number ranging from $-$1.5 to 0.29. Dispersion measurements were carried using a fast response flame ionisation detector.
	The results show that the stratification effect on the plume width is significantly lower than the effect on the vertical profiles. Stable stratification did not affect the plume central axis inside the canopy, but in the unstable case the axis appeared to deviate from the neutral case direction. Above the canopy both stratification types caused an increase in the plume deflection angle compared to the neutral case. Measured mean concentrations in stable stratification were up to two times larger in the canopy compared to the neutral case, while in convective conditions they were to three times smaller. The proportionality between the vertical turbulent fluxes and the vertical mean concentration gradient was also confirmed in the stratified cases.
	The high-quality experimental data produced during this work may help developing new mathematical models and parametrisation for non-neutral stratified conditions, as well as validating existing and future numerical simulations.
\end{abstract}




\end{frontmatter}


\section{Introduction}

Atmospheric stratification can have a significant impact on pollutant dispersion in urban areas, but there are still many uncertainties in quantifying its effect, mainly because of the difficulties of studying non-neutral conditions in the laboratory and the field. Urban dispersion models generally discard stratification effects based on the fact that in cities, due to their large aerodynamic roughness length, mechanically-generated turbulence tend to dominate over buoyancy effects \citep{Britter2003}. This seems a sensible assumption, but it is largely unsupported by observations. \citet{Wood2010}, for example, found that either stable or convective conditions represent a large majority of cases in a large urban area.

Nevertheless, laboratory studies in non-neutrally stratified conditions are very rare, especially when dealing with large urban building arrays. The case of stable and unstable incoming flow over either an aligned or staggered array of cubes has been investigated by \citet{Uehara2000} and \citet{Kanda2016}, respectively. The former focused on a cross section downstream a block, with just one vertical profile scanning the entire boundary layer depth. Moreover, neither heat fluxes nor pollutant concentration measurements were attempted. \citet{Kanda2016} expanded further with measurements of heat fluxes and mean concentration for a point source release, but only one full-height vertical profile was acquired and no concentration fluctuations and fluxes were sampled. Moreover, only one stable and one unstable cases were considered. The concentration and turbulence measurements in and above the canopy revealed important effects
of the stratification, encouraging further studies in this direction. In particular, the plume depth and width were affected by stratification, being both smaller in the SBL case and larger in the CBL one, compared to the NBL reference.

Slightly more abundant are the numerical studies, especially involving large eddy simulations \citep[LES,][]{Inagaki2012,Park2013,Xie2013,Boppana2014}.
\citet{Tomas2016} simulated the effect of stable stratification on flow and dispersion from a line source over an array of aligned cubes. They found that under a weak SBL (bulk Richardson number based on the boundary-layer depth, $Ri_\delta=0.15$) the depth of the internal boundary layer (IBL) after 24 rows of cubes was 14\% shallower compared to a NBL, while the turbulent kinetic energy (TKE) was reduced by 21\%. On the other hand, the area-averaged street concentration level in a SBL was found to be 17\% larger than for the NBL thanks to the decreased streamwise advection and pollutant trapping in the IBL.

\citet{Shen2017} simulated a SBL developing over an array of aligned cubes. Their model was validated using results from \citet{Kanda2016}. Different plan area densities ($\lambda_p$) were investigated, ranging from isolated roughness to skimming flow regimes. A point-source ground-level pollutant release was also considered. Results showed that the reduced advection velocity in the SBL is the cause for the larger concentration in the canopy.
\citet{Jiang2018} employed the same array of aligned cubes but with a weaker CBL case (bulk Richardson number based on the cubes' height, $Ri_H = -0.15$) and a line source. Results showed that a primary recirculation region was formed inside the canopy, similar to the one observed in bi-dimensional street canyons \citep[see, e.g.,][in this regard]{Cheng2011}. The turbulent pollutant fluxes were found to considerably contribute to the pollutant transport into the ``canyon'', especially in the side ends of the streets, while no inflow due to turbulence was detected vertically from the top section. On the other end, turbulent fluxes were found to be the main contributor for pollutant going out of the ``canyon'' from the top surface.

The work presented in this paper is part of the StratEnFlo project, funded by the UK Engineering and Physical Research Council (EPSRC). It was a first attempt to bridge the identified gap in the literature about the lack of experimental data in non-neutral conditions. Initially, new methodologies were developed and optimised to simulate either stable or convective conditions in a meteorological wind tunnel, producing a boundary layer that was thick enough for urban studies \citep{Marucci2018}. The non-neutral boundary layers produced in that first phase were then applied to a single heated/cooled street canyon \citep{Marucci2019} and to an array of rectangular buildings \citep{Marucci2019flow}. The latter, in particular, studied the effects of several incoming SBLs and CBLs on the flow over and within the urban array (using a wind direction of 45$^\circ$), finding that the modifications on the flow and turbulence fields caused by even the weak stratification levels tested were significant. The experiments designed by \citet{Marucci2019flow} also included dispersion measurements, but results were not discussed in that manuscript.

\cite{Sessa2018,Sessa2019} employed the dataset produced in the present study (but with 0$^\circ$ wind direction) to validate their LES simulation for a rectangular array of buildings with different levels of SBL (ranging from $Ri_\delta$ 0.21 to 1.0). Pollutant release from either a linear or a point source was also modelled. Mean velocity, Reynolds stresses and mean concentrations were in good agreement with the wind tunnel experiments. The mean concentration below the canopy in case of line source for $Ri_H = 1$ was twice as large as the one for $Ri_H = 0.2$ , while for the same stratification cases the concentration from the point source was four times larger. This was partially attributed to simultaneous decrease of both lateral and vertical scalar spreading in the case of point source release. The vertical turbulent fluxes from the line source release in several streamwise locations confirmed the decrease of the vertical scalar mixing for increasing stratification. They also observed a reduction with increasing stratification of the height where the vertical flux became negligible.

This paper reports the results of the dispersion experiments mentioned by \citet{Marucci2019flow}, with a detailed analysis of the tracer concentration measurements and a discussion on their significance in terms of urban pollution. Section \ref{sec:methods} describes the employed facilities and the experimental settings, as well as the urban model used for this study. The flow characteristics and approaching flow conditions, reported in detail by \cite{Marucci2019flow}, are summarised in section \ref{sec:flow}. Results and discussion about the plume characteristics are reported in section \ref{sec:plume}, while section \ref{sec:flux} analyses the mass flux results in more details. Conclusions are reported in section \ref{sec:Conclusion}.

\section{Experimental methodology}
\label{sec:methods}
The EnFlo meteorological wind tunnel at the University of Surrey is an open-circuit suction boundary-layer wind tunnel with a working section size of 20~m$\times$3.5~m$\times$1.5~m. A turbulent boundary layer was generated using two sets of Irwin spires \citep{Irwin1981}, one for the SBL study and one for the CBL, and roughness elements covering the floor upstream of the model \citep[see, e.g.][for more details]{Marucci2018,Marucci2019flow}. A vertical inlet temperature profile can be imposed when working in stratified conditions and the wind tunnel floor can be either cooled or heated depending on the atmospheric conditions to be studied. The optimised techniques to generate either stable or convective boundary layers in this wind tunnel have been fully described by \citet{Marucci2018}.

The nominal reference velocity ($U_{REF}$) was used as a target for the closed-loop system controlling the two fans at the outlet of the wind tunnel, based on the measurements by an ultrasonic anemometer placed 5~m downstream of the inlet section, 1~m from the wind tunnel centre line (laterally) and 1~m high. The coordinate system used in this paper is aligned with the urban array model, originating at the centre of the wind tunnel turntable (14~m downstream of the inlet). When the wind direction was set to 0$^\circ$ the $x$-axis was aligned with the tunnel centre line, the $y$-axis was in the lateral direction and the $z$-axis was the vertical one.

The model used in this study was originally developed for the DIPLOS project \citep[see][]{Castro2017,Fuka2018,Hertwig2018} and includes more than 350 rectangular blocks with dimensions $H\times2H\times H$ (width$\times$length$\times$height) regularly spaced (spacing $H = 70$~mm). This geometry is regular, yet is more complex than the classical cubical array and typical street canyon features start to show up \citep{Castro2017}, especially in non aligned configurations (i.e.~when the wind direction is not aligned with the streets). For this reason all the experiments reported here were carried out using a 45$^\circ$ model rotation. In order to validate LES numerical results \citep{Sessa2018}, the data set also includes some experiments with 0$^\circ$, but results are not reported here.

\begin{figure}
	\centering
	\includegraphics[width=\linewidth]{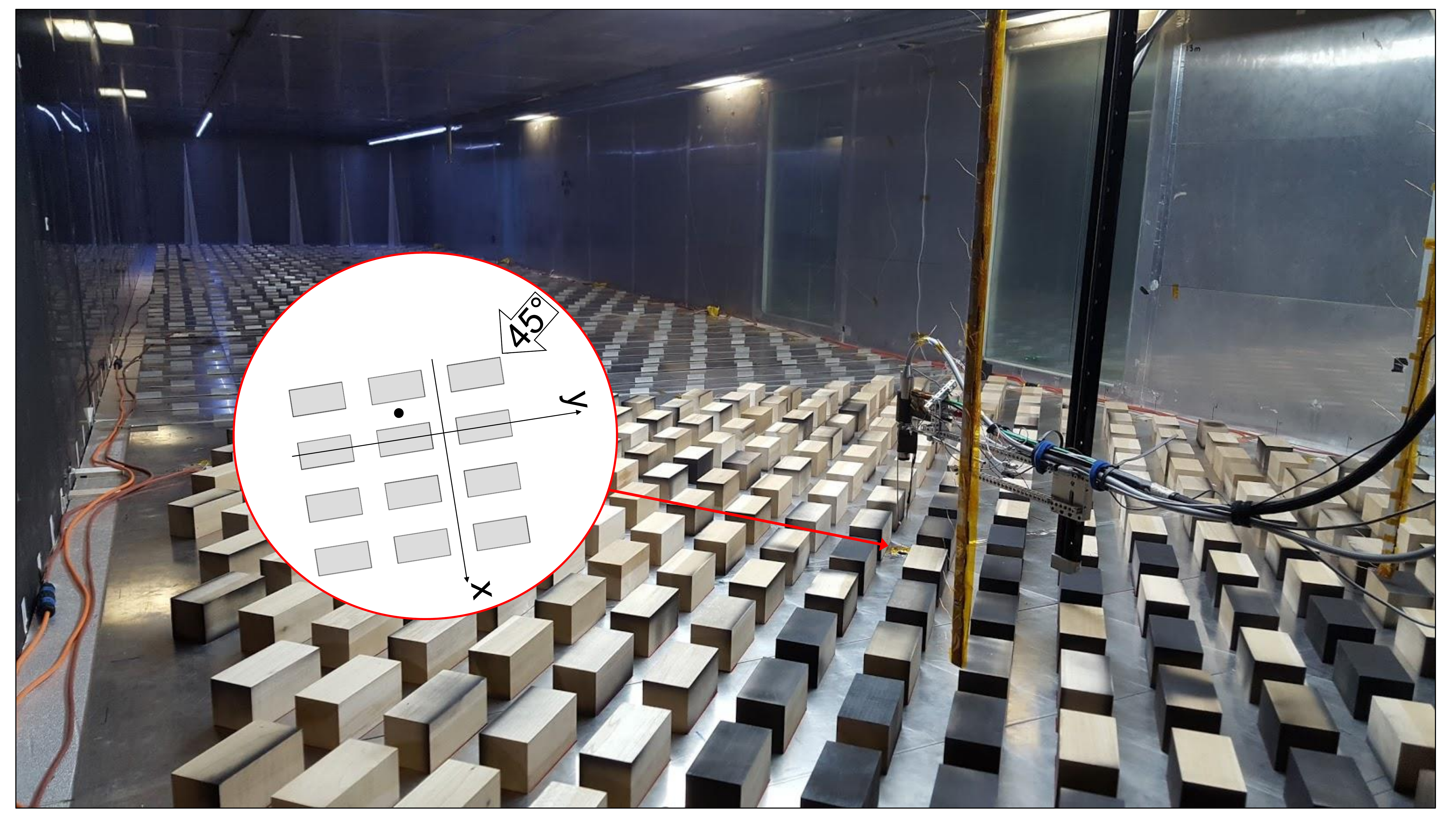}
	\caption{Urban array in the wind tunnel and schematics of the model centre. The source location is indicated by the black dot.}
	\label{fig:UrbanArrayModel}
\end{figure}

In Fig.~\ref{fig:UrbanArrayModel} a photo and a schematic of the employed urban array model are displayed. Al the experiments reported here were performed using a wind direction of 45 degrees. Dispersion experiments were carried out by using a tracer gas released from a circular source (diameter 22~mm) located at ground level at the centre of the street canyon created by the long edge of a building close to the centre of the model. The tracer was a mixture of propane (not exceeding 1.8\%) in air with an exit velocity maintained low, at $0.03U_{REF}$, in order to simulate a passive emission.

The measurement setup is described in a detailed manner by \citet{Marucci2018}, \citet{Marucci2019} and \citet{Marucci2019flow}. Temperatures, concentrations and two components of velocity were measured simultaneously using, respectively, a fast-response cold-wire probe (CW), a fast-response flame ionisation detector (FFID) and a laser Doppler anemometer (LDA). The LDA target acquisition frequency was set to 100~Hz, while both temperatures and concentrations were sampled at 1000~Hz. Given the irregular nature of the LDA measurements and the different frequencies, a resampling and synchronisation of the three signals was necessary for computing heat and mass fluxes \citep{Marucci2019}.

Each measurement point was sampled for 2.5 minutes, following previous experiments in neutral \citep{Castro2017} and non-neutral \citep{Marucci2019} conditions. The standard errors for first and second order statistics was evaluated at each measurement point and deemed satisfactory for high-quality experiments \citep[see, in particular,][]{Marucci2019flow}. As far as concentration measurements are concerned, in stable conditions standard errors for mean concentrations ($\overline{C}$) were below 10\%, while variance ($\overline{c'^2}$) values were generally 20\%. Standard errors were, as expected, higher for neutral and convective conditions, suggesting that longer averaging times might be needed for the CBL cases in future experiments. Standard errors for covariance values ($\overline{u'c'}$, $\overline{v'c'}$ and $\overline{w'c'}$) were generally between 10 and 25\%, with little sensitivity to different stratification conditions. In the previous discussion and throughout the paper, capital letters and overbars represent a time averaged value, while small letters and the prime symbol identify fluctuating components.

\section{Approaching flow and boundary layer over the array}
\label{sec:flow}
Five different non-neutral boundary layers were generated in this study (3 SBLs and 2 CBLs), and they were compared with two neutral reference cases. Two NBLs were required as the non-stratified cases were reproduced using two sets of spires, matching the ones used in the corresponding stratified case (one for stable flows and one for convective). The different heights used for the spires is the main reason why some of the quantities in the reference neutral cases differ from each other. The measured and nominal properties in the five cases are summarised in Tab.~\ref{table:aerpar}. The nominal Richardson number for each experiment ($Ri_\delta^{app}$) is the desired value in the approach flow, which sometimes differ slightly from the actual value measured over the array ($Ri_\delta$, also reported in the table. The two types of bulk Richardson numbers used in this paper ($Ri_\delta$ and $Ri_H$) can be calculated as

\begin{equation} \label{bulkRi}
Ri_\delta = \frac{g\left(\Theta_\delta - \Theta_0\right)\delta}{\Theta_0 U_\delta^2}, \ \
Ri_H = \frac{g\left(\Theta_H - \Theta_0\right)H}{\Theta_0 U_H^2}
\end{equation}

\noindent where $\Theta$ symbols represent temperatures, $U$ velocities, the subscripts $\delta$ and $H$, respectively, the boundary-layer depth and the buildings' height, $g$ is the gravitational acceleration and $\Theta_0$ is a reference temperature measured close to the floor (at $z=10$~mm).

\begin{table}
	\caption{Nominal and measured parameters for the two sets of experimental cases (stable and convective). Two different neutral reference cases ($Ri_\delta^{app}=0$) are reported as the values differ slightly due to the different sets of spires used.}
	\centering
	\label{table:aerpar}
	\scalebox{0.75}{
		\begin{tabular}{l c c c c | c c c}
			\toprule
			& \multicolumn{4}{c|}{\textbf{SBL cases}} & \multicolumn{3}{c}{\textbf{CBL cases}} \\
			\midrule
			$\mathrm{Ri_\delta^{app}}$ & 0 & 0.14 & 0.21 & 0.29 & 0 & $-$0.5 & $-$1.5 \\
			$\mathrm{\Delta\Theta_{\text{MAX}}\ (^\circ C)}$ & 0 & 10.8 & 16 & 17.8 & 0 & $-$24.2 & $-$39.2\\
			$\mathrm{U_{\text{REF}}\ (m/s)}$ & 1.25 & 1.25 & 1.25 & 1.15 & 1.25 & 1.25 & 1.0\\
			\midrule
			$\mathrm{u_\ast/U_{\text{REF}}}$ & 0.078 & 0.063 & 0.061 & 0.059 & 0.081 & 0.105 & 0.118\\
			$\mathrm{z_0\ (mm)}$ & 3.45 & 2.5 & 2.6 & 2.9 & 4.0 & 6.3 & 6.2\\
			$\mathrm{d\ (mm)}$ & 52.5 & 53.5 & 54.5 & 55.0 & 50.8 & 23.5 & 21.5\\
			$\mathrm{\delta\ (mm)}$ & 850 & 850 & 850 & 850 & 1000 & 1200 & 1350\\
			$\mathrm{\Theta_0\ (^\circ C)}$ & - & 17.4 & 17.8 & 18 & - & 39.0 & 50.0\\
			$\mathrm{\Delta\Theta \left[=\Theta_{\delta}-\Theta_0\right]}$ & - & 8.2 & 12.8 & 14.3 & - & $-$15.8 & $-$24.6\\
			$\mathrm{\theta_\ast\ (^\circ C)}$ & - & 0.221 & 0.315 & 0.355 & - & $-$0.60 & $-$0.92\\
			$\mathrm{w_\ast/U_{REF}}$ &-& - & - & - & - & 0.115 & 0.158\\
			$\mathrm{z_{0h}\ (mm) \left[d_h=d\right]}$ & - & 0.006 & 0.004 & 0.006 & - & 0.0067 & 0.0037\\
			$\mathrm{d_h\ (mm) \left[Fitted\right]}$ & - & 51.4 & 47.3 & 37.4 & - & 52.3 & 44.5\\
			$\mathrm{z_{0h}\ (mm) \left[d_h\ fitted\right]}$ & - & 0.006 & 0.004 & 0.010 & - & 0.0050 & 0.0030\\
			\midrule
			$\mathrm{\delta/L}$ & 0 & 0.40 & 0.62 & 0.88 & 0 & $-$0.51 & $-$1.09\\
			$\mathrm{u_\ast/w_\ast}$ & - & - & - & - & - & 0.92 & 0.75\\
			$\mathrm{Ri_\delta}$ & 0 & 0.12 & 0.19 & 0.24 & 0 & $-$0.35 & $-$0.91\\
			$\mathrm{Ri_H}$ & 0 & 0.10 & 0.19 & 0.28 & 0 & $-$0.15 & $-$0.19\\
			$\mathrm{Re_\ast}$ & 22.7 & 11.2 & 13.3 & 11.8 & 26.8 & 49.5 & 40.8\\
			$\mathrm{Re_\delta\ \left(x10^3\right)}$ & 67.09 & 78.57 & 79.95 & 74.30 & 87.8 & 92.7 & 74.6\\
			\bottomrule
	\end{tabular}}
\end{table}

Stable boundary layers were generated by imposing a non-uniform inlet temperature profile, cooling the floor at a desired temperature and adjusting the maximum inlet temperature ($\Delta\Theta_{MAX}$ is defined as the difference between this maximum temperature and the floor temperature) and reference velocity ($U_{REF}$) to set the required stratification strength \citep{Marucci2018}. It should be noted that $Ri_\delta^{app}$ in the table is the nominal (or desired) bulk Richardson number of the approaching flow, which sometimes differs slightly from the one actually measured (also reported in the table). Convective boundary layers were generated by setting a uniform inlet temperature profile capped by a linear inversion of roughly 10$^\circ$~C/m starting from 1~m upwards, heating the floor using an optimised layout for the heating panel mats and adjusting $\Delta\Theta_{MAX}$ and $U_{REF}$ \citep{Marucci2018}.

Surface aerodynamic (friction velocity $u_\ast$, roughness length $z_0$, displacement height $d$, BL detpth $\delta$) and thermal (scaling temperature $\theta_\ast=-\left( \overline{w'\theta'} \right)_0 / u_\ast$, thermal roughness length $z_{0h}$, thermal displacement height $d_h$) were estimated as described in details by \citet{Marucci2019} and \citet{Marucci2019flow}, by fitting the logarithmic profiles and the vertical shear stress profiles. Other values reported in the table are a reference temperature close to the floor ($\Theta_0$), the temperature at the boundary-layer height ($\Theta_\delta$), a velocity scale valid on the mixed layer of a CBL, defined as \citep{Kaimal1994}:

\begin{equation}
 w_\ast = \left[ \frac{g}{\Theta_0}\left(\overline{w'\theta'}\right)_0 \delta\right]^{1/3}
\end{equation}

\noindent the Monin-Obukhov length ($L$), the bulk Richardson numbers measured at the boundary-layer depth ($Ri_\delta$) and building height ($Ri_H$), the Reynolds number ($Re_\delta$) and roughness Reynolds number ($Re_\ast$).

A full analysis of the boundary layer flow, turbulence and temperature fields over the urban array in the five stratification cases considered here is reported by \citet{Marucci2019flow}.

\section{Plume characteristics}
\label{sec:plume}
\subsection{Stable stratification}
\begin{figure}
	\centering
	\includegraphics[width=\linewidth]{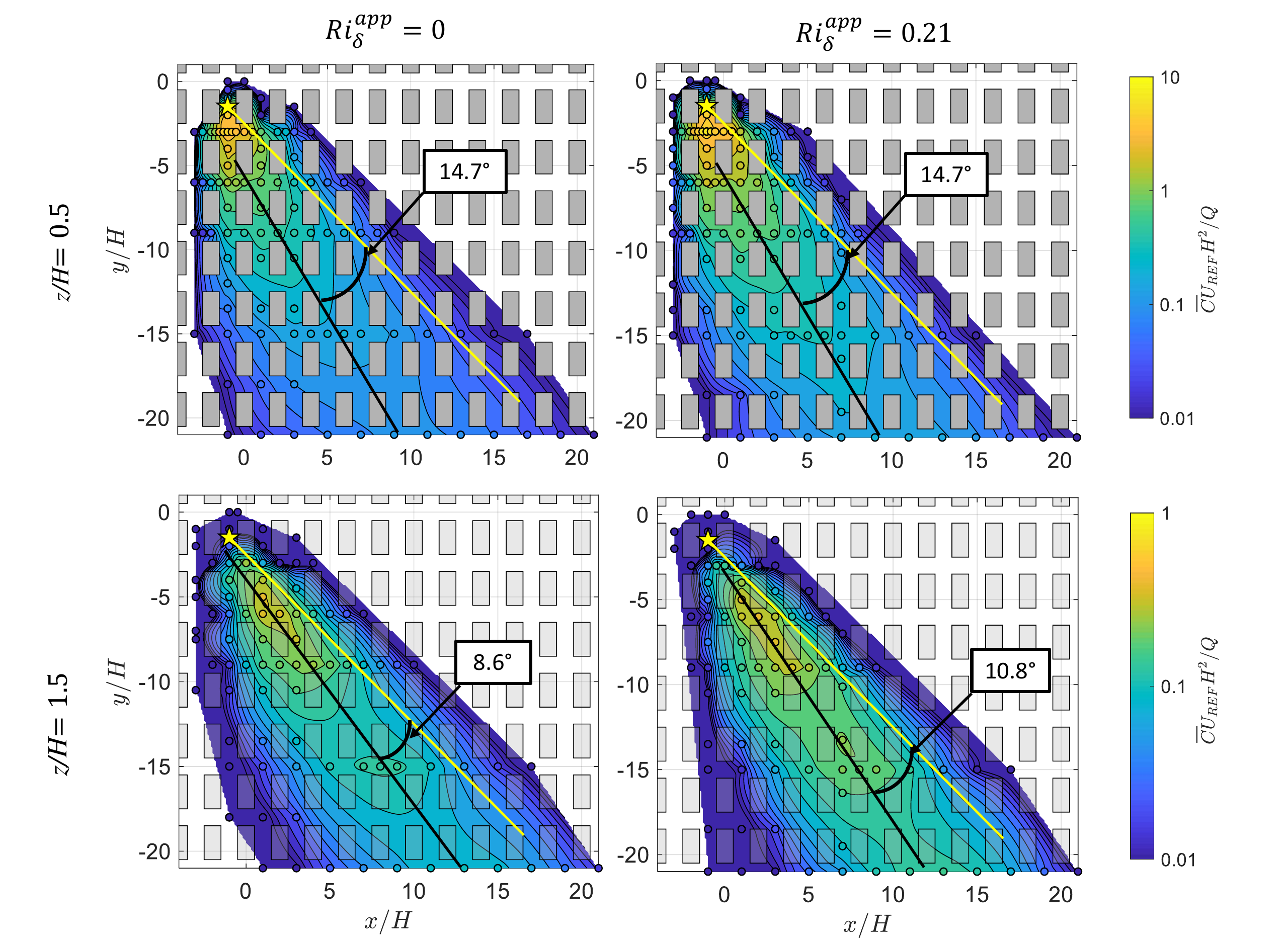}
	\caption{Contour plots of non-dimensional mean concentration for NBL and SBL inside and above the canopy for wind direction 45$^\circ$. Black line is plume centreline, yellow line is free-stream wind direction.}
	\label{fig:ConcSBLcont}
\end{figure}

In Fig.~\ref{fig:ConcSBLcont} contour plots of pollutant mean concentration are shown for the NBL and a SBL case ($Ri_\delta^{app}  = 0.21$) both inside ($z/H = 0.5$) and above ($z/H = 1.5$) the urban canopy. The tracer was released from a ground level source located at $x/H = -1$ and $y/H = -1.5$. The plume central axis -- defined as the straight line that minimises the distance from the mean values in the Gaussian fit of the lateral profiles (see equation \ref{gaussianCurveLat}) -- does not seem to be affected by the stable stratification inside the canopy. As a matter of fact, its axis appears to deviate from the free-stream wind direction due to channelling effect by about 14.7$^\circ$ both in neutral and stable atmospheric conditions. The channelling is caused by the presence of the small street canyons and it is even more evident in the first $2H$ downstream of the source, where the plume axis is almost coincident with the long street centreline. Above the canopy the plume axis still presents a deflection from the free-stream wind direction, despite the fact that the flow field is already completely aligned with the tunnel axis \citep{Marucci2019flow}. The angles are slightly different, though (8.6$^\circ$ for NBL and 10.8$^\circ$ for SBL). Since the actual wind direction is already aligned with $45^\circ$ at $z/H = 1.5$ and above, the different plume angle is just a result of the different distribution of concentrations closer to the ground. In facts, pollutant concentrations in the canopy remain larger further away from the source in case of stable stratification. It would be interesting to compare these results to cases with a different Richardson number, but unfortunately we do not have enough data to estimate the plume direction for other stratification levels.

\begin{figure}
	\centering
	\includegraphics[width=0.9\linewidth]{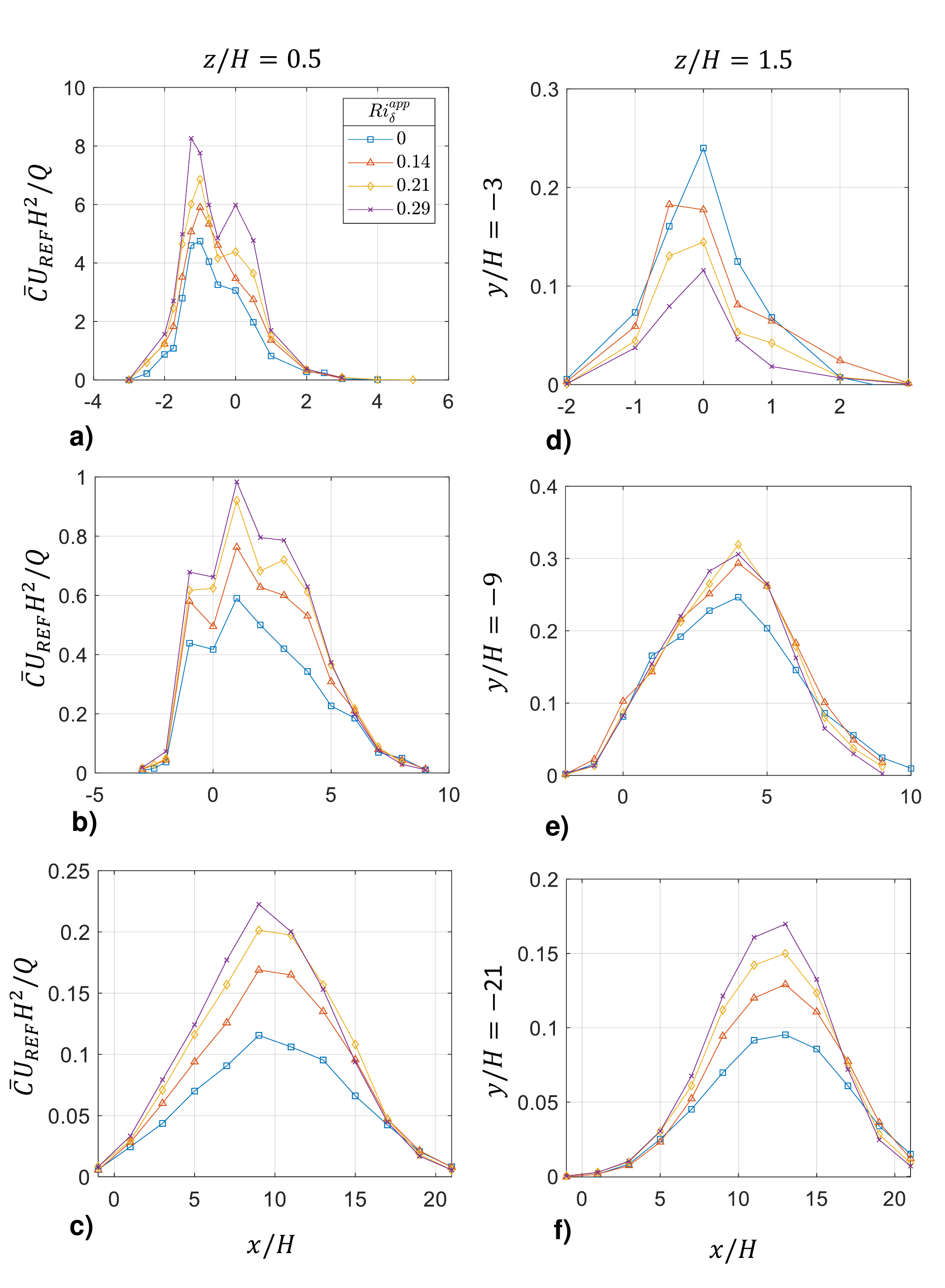}
	\caption{Lateral profiles of mean concentration inside and above the canopy for four levels of stability.}
	\label{fig:LateralPlumeSBL}
\end{figure}

The plume width does not appear to be significantly affected by the applied stratification inside the canopy, with just a small reduction. A similar statement can be made for the plume above. This can be better appreciated from the lateral profiles of mean concentration shown in Fig.~\ref{fig:LateralPlumeSBL}, where the values for two other levels of stability are plotted as well.

The mean concentration values, on the contrary, show a clear effect of the different stratification levels. In all the graphs shown in Fig.~\ref{fig:LateralPlumeSBL}, the concentration – both inside and immediately above the canopy – appears larger in the SBL and increasing with $Ri_\delta$ up to about twice as large. The only exception is in the upper region closer to the source, in which the trend is inverted. This behaviour is expected and due to the reduced vertical displacement of the flow under a SBL.

\begin{figure}
	\centering
	\includegraphics[width=1\linewidth]{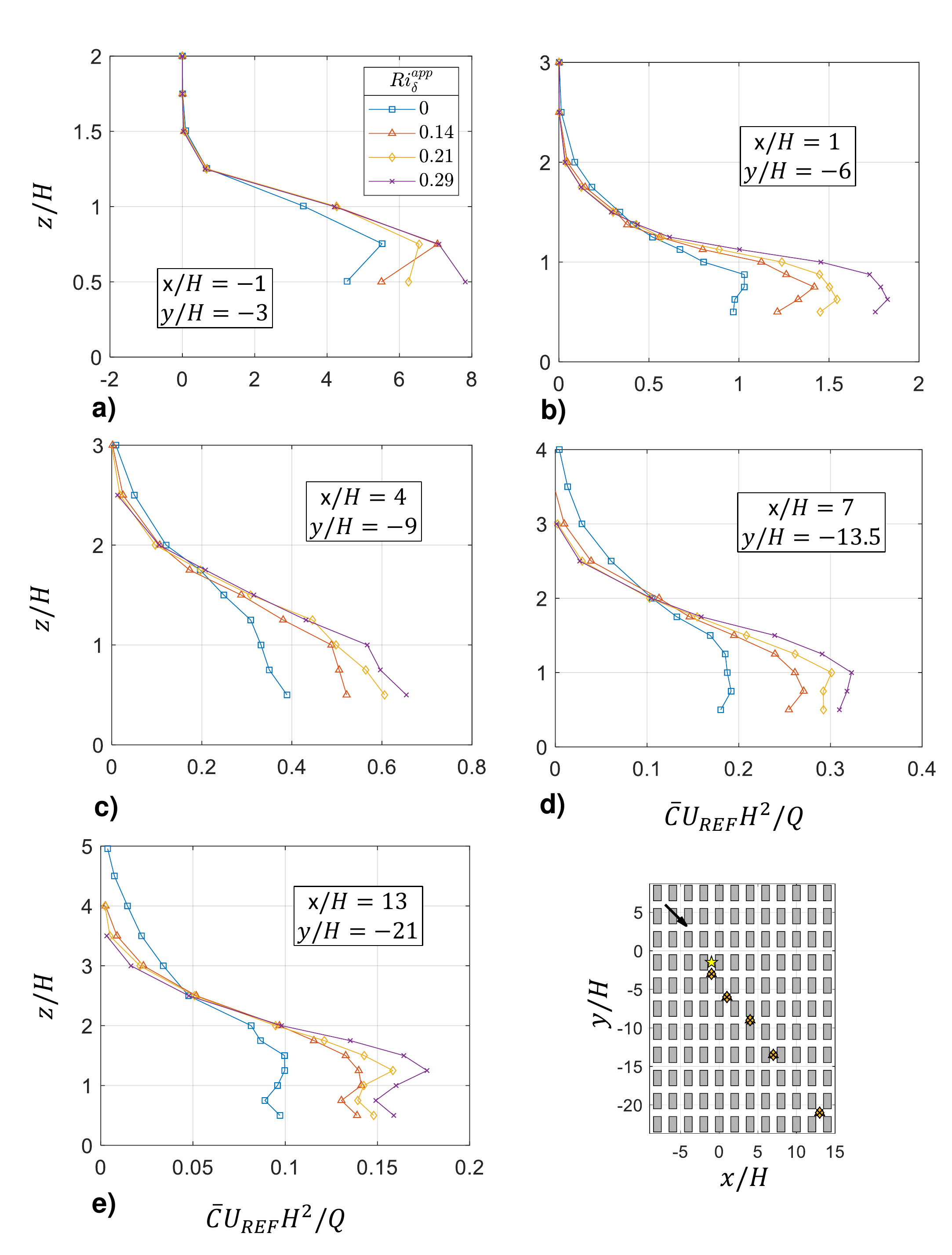}
	\caption{Vertical profiles of mean concentration approximately along the plume axis for four levels of stability. The star on the map at bottom-right corresponds to the source, while the other marks show the locations of the five vertical profiles.}
	\label{fig:VerticalPlumeSBL}
\end{figure}

The plume vertical depth is smaller under stable stratification, as shown in Fig.~\ref{fig:VerticalPlumeSBL}. It is also possible to note how all the SBL cases seem to behave similarly above 1.5$H$, showing the same plume depth reduction of up to 30\% compared to the NBL. Within the canopy, the concentration level appears approximately constant with height, at least down to the lowest measured position (0.5$H$). All measured profiles show a similar behaviour with different levels of stratification (Fig.~\ref{fig:VerticalPlumeSBL}), confirming that the modification induced by the stable boundary layer are independent from the particular location within the urban array. The chosen positions are indeed different in terms of mixing properties, with three of them at street intersections, one in a ``long'' street canyon and one in a ``short'' street canyon, yet the changes due to different levels of stratification seem to apply to all of them in a similar way.

\begin{figure}
	\centering
	\includegraphics[width=0.6\linewidth]{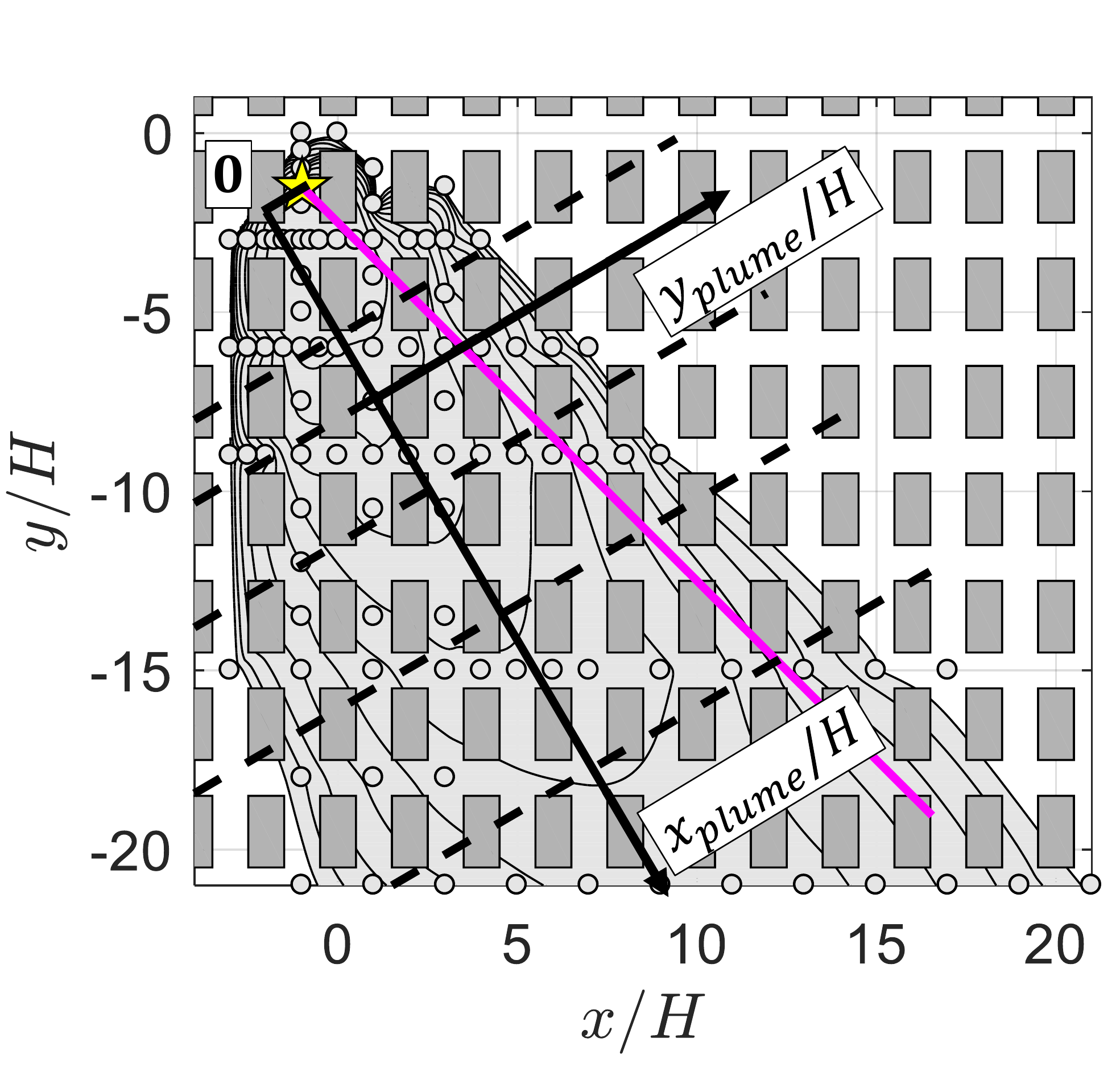}
	\caption{Plume axes reference system.}
	\label{fig:PlumeAxisScheme}
\end{figure}

\begin{figure}
	\centering
	\includegraphics[width=1\linewidth]{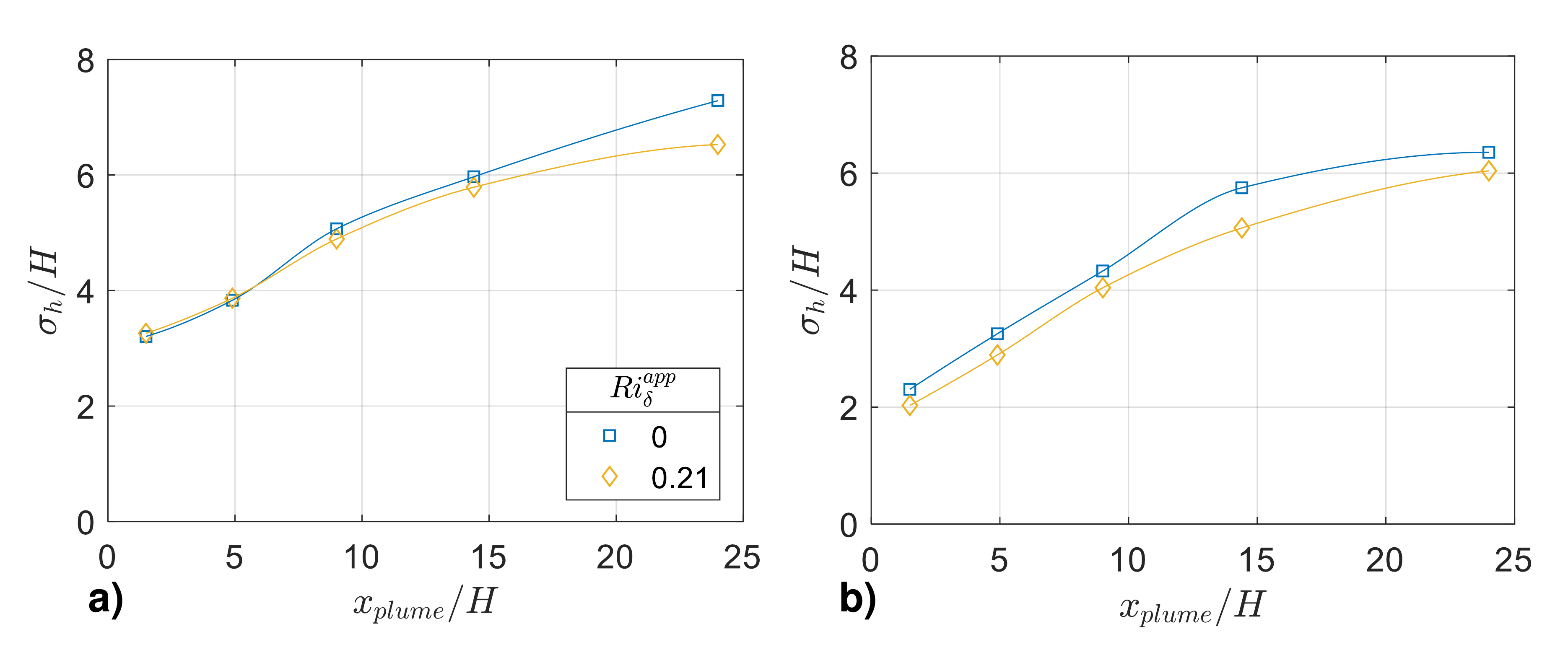}
	\caption{$\sigma_h$ for SBL and NBL varying the distance from the source at $z/H$ of 0.5 (a) and 1.5 (b).}
	\label{fig:SigmahSBL}
\end{figure}

\begin{figure}
	\centering
	\includegraphics[width=0.6\linewidth]{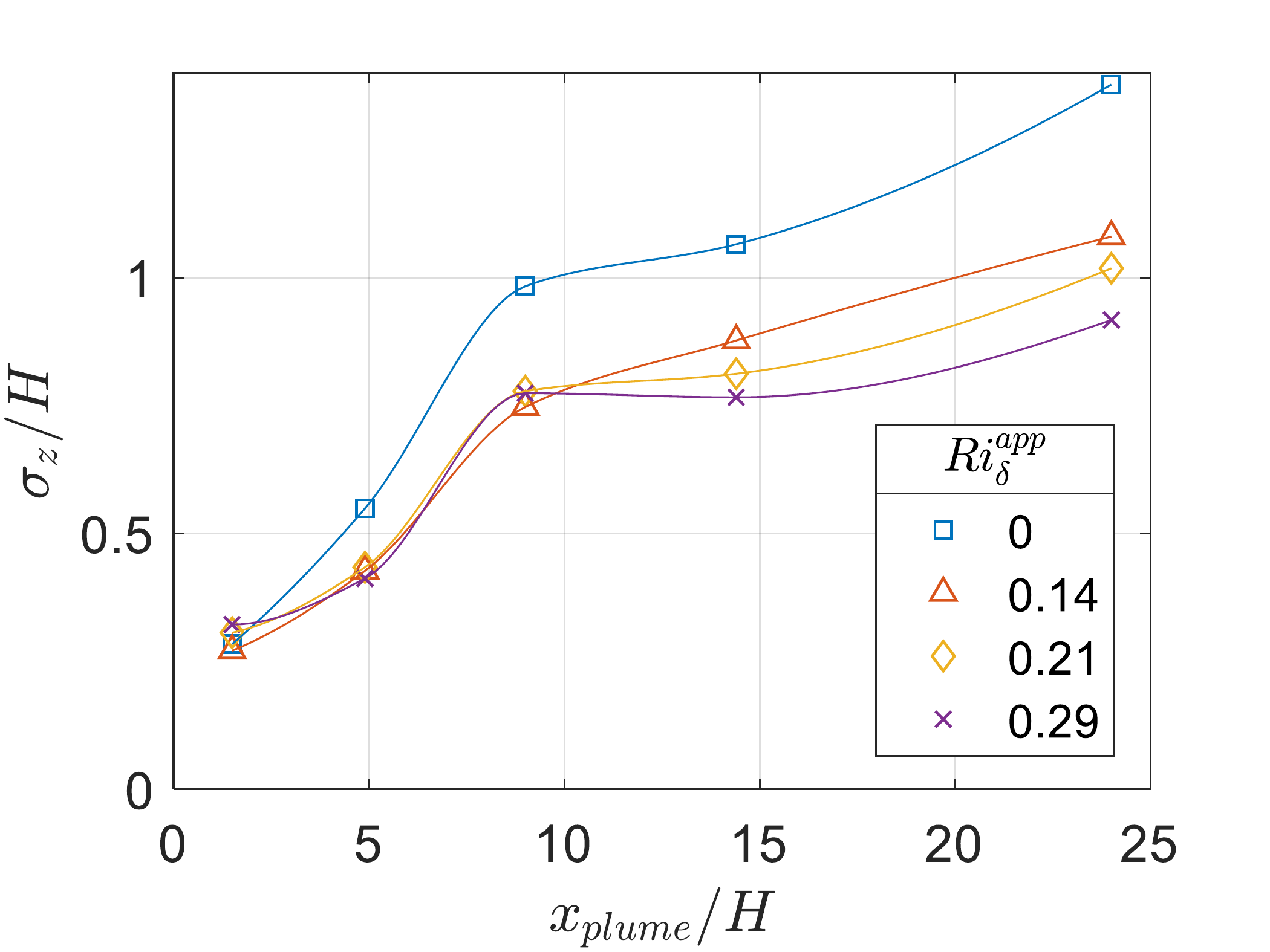}
	\caption{$\sigma_z$ for SBL and NBL varying the distance from the source.}
	\label{fig:SigmazSBL}
\end{figure}

In order to better quantify the effect on the width and depth of the plume, a fitting was attempted with a Gaussian distribution. The following curve

\begin{equation}\label{gaussianCurveLat}
\overline{C} = Ae^{-\frac{(y_{plume}-\mu)^2}{2\sigma_h^2}}
\end{equation}

\noindent in which $A$, $\mu$ and $\sigma_h$ are free fitting parameters, was fitted by means of a non-linear least squares method to profiles extrapolated from the contour plots, perpendicular to the axis of the plume indicated in Fig.~\ref{fig:ConcSBLcont}. On this regard, two axes were defined, $x_{plume}$ which coincides with the plume axis, and $y_{plume}$, perpendicular to the former, as shown in Fig.~\ref{fig:PlumeAxisScheme}. The Gaussian fit was remarkably satisfactory for all measurement profiles, at all distances from the source. In Fig.~\ref{fig:SigmahSBL} the values obtained for $\sigma_h$ (representative of the plume width along $y_{plume}$) are displayed for the neutral reference and the $Ri_{\delta}^{app} = 0.21$ case for five $x_{plume}$ locations (the origin of the plume reference system was chosen so that $x_{plume}$ represented the distance of the lateral profiles from the source). The trend of $\sigma_h$ shows that inside the canopy the plume width is only very slightly reduced by the stable stratification, and only far from the source. Above, instead, a difference (but still very small) is discernible throughout the plume.

The $\sigma_z$ plot (Fig.~\ref{fig:SigmazSBL}) -- obtained using the Gaussian fit on a similar equation as Eq.~\ref{gaussianCurveLat}, but with $\sigma_h$ replaced by $\sigma_z$ and $y_{plume}$ by $z$ -- confirms that the plume depth is very similar in the three considered stability cases, starting to differ only after 10$H$ from the source. It is possible to note that the values of $\sigma_z$ appeared to be more sensitive to the stable stratification than $\sigma_h$. This is in agreement with what observed by \cite{Briggs1973} in field experiments over urban roughness. On the contrary, \cite{Kanda2016} found the plume depth only slightly affected, while the width was sensibly reduced by the application of the stable stratification. A complete explanation of this peculiar behaviour was not given, but \citet{Kanda2016} mentioned possible uncertainties due to small variations in depth and width.

\begin{figure}
	\centering
	\includegraphics[width=1\linewidth]{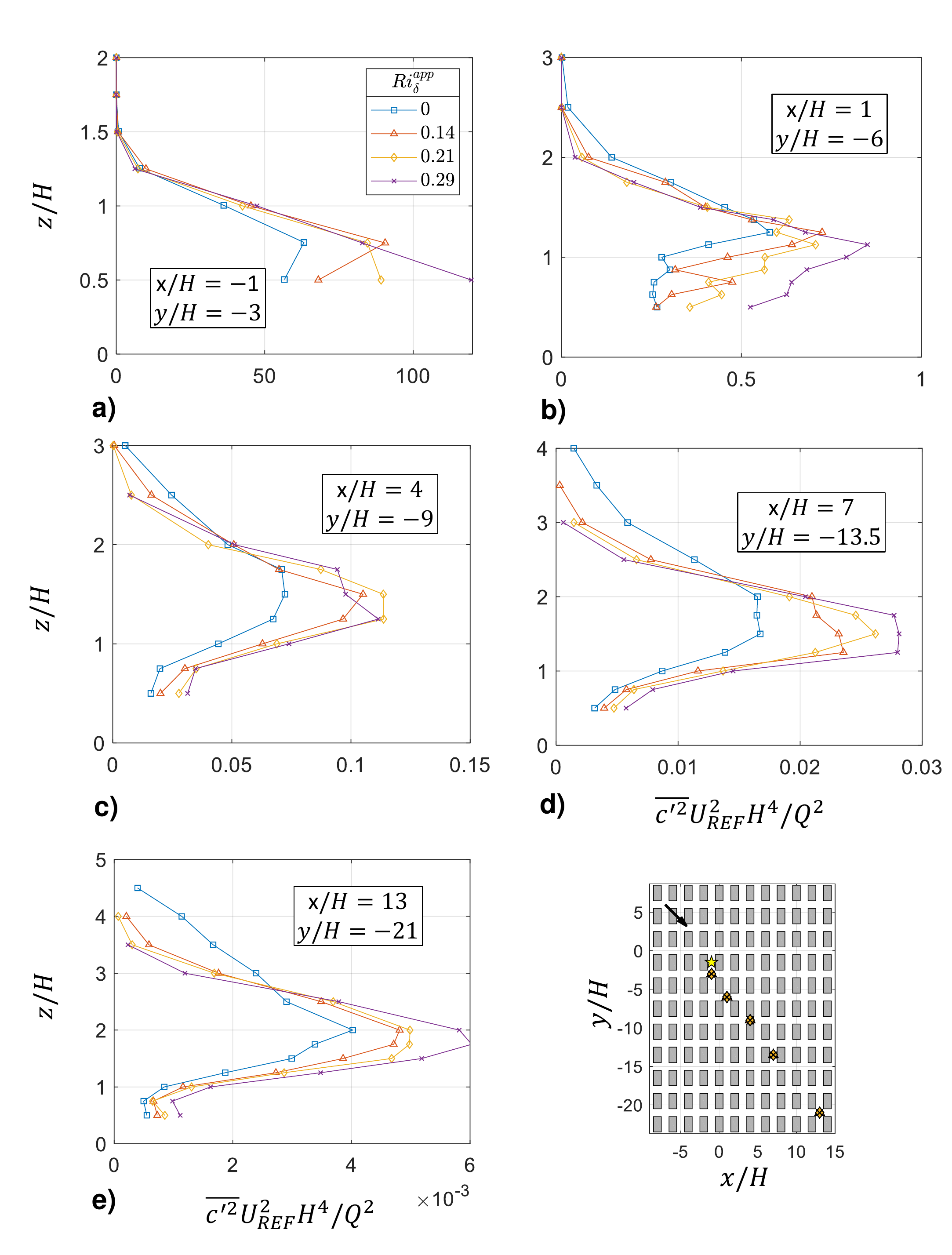}
	\caption{Vertical profiles of concentration variance approximately along the plume axis for four levels of stability. The star on the map at bottom-right corresponds to the source, while the other marks show the locations of the five vertical profiles.}
	\label{fig:VerticalPlumevarianceSBL}
\end{figure}

The lateral concentration fluctuation profiles at 0.5 and 1.5$H$ (Fig.~\ref{fig:VerticalPlumevarianceSBL}) have a similar trend to the mean concentration, varying with stratification in the same manner. The behaviour of the vertical profile, though, is different up to $z/H = 2$, where the fluctuations present an increase to a maximum above the canopy, followed by a reduction further above. Nevertheless, the amplification or reduction of the variance values following the stratification is similar to what experienced by the mean concentrations.

\subsection{Unstable stratification}
\begin{figure}
	\centering
	\includegraphics[width=\linewidth]{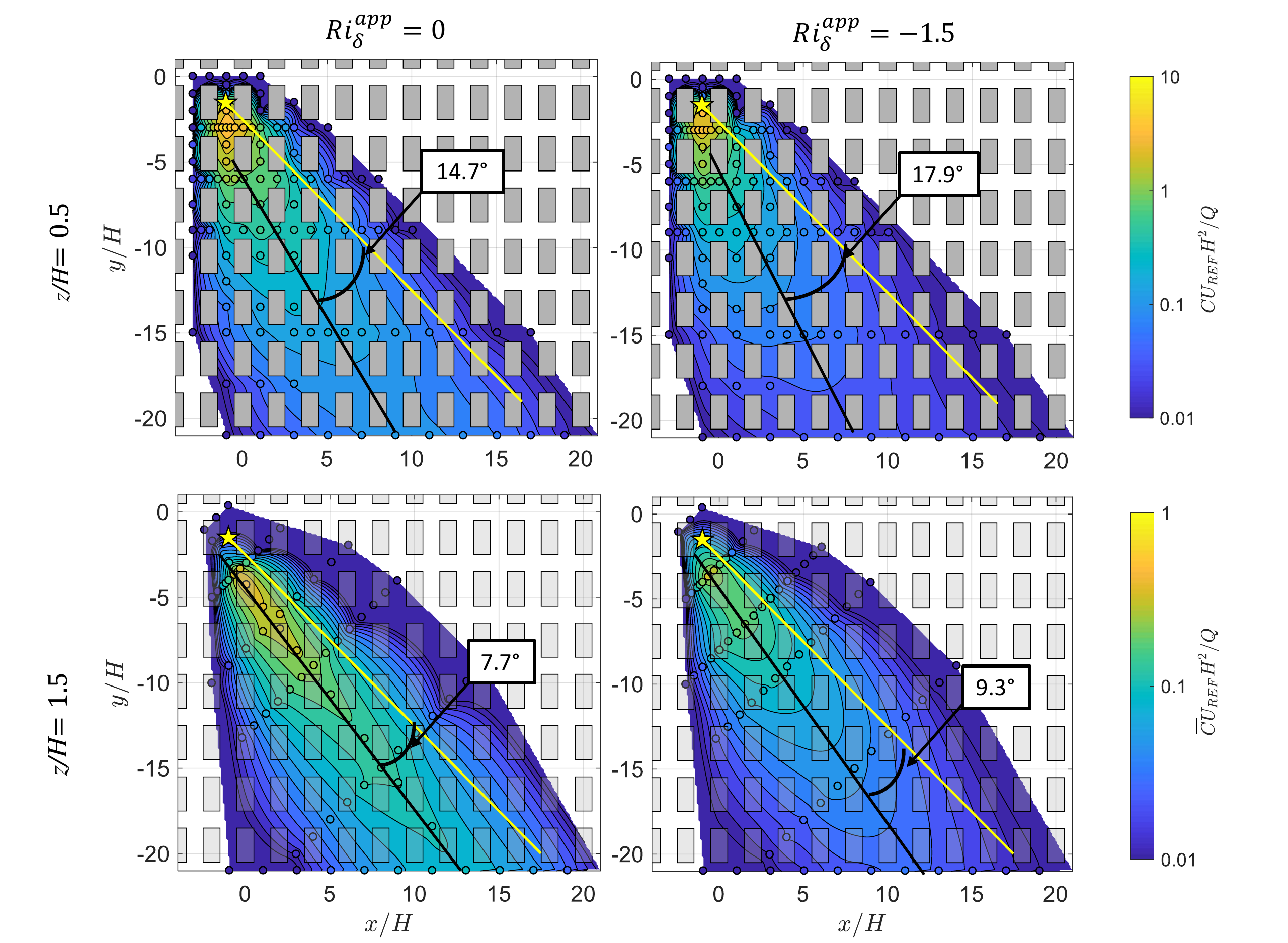}
	\caption{Contour plots of non-dimensional mean concentration for NBL and CBL inside and above the canopy for wind direction 45$^\circ$. Black line is plume centreline, yellow line is free-stream wind direction.}
	\label{fig:ConcCBLcont}
\end{figure}

Fig.~\ref{fig:ConcCBLcont} shows contour plots of pollutant mean concentration for the NBL and a CBL case ($Ri_{\delta}^{app}  = -1.5$) both inside ($z/H = 0.5$) and above ($z/H = 1.5$) the canopy. The same source location as for the stable cases has been used ($x/H = -1$, $y/H = -1.5$). Differently from the considered SBL cases, the plume central axis here appears modified by the unstable stratification also inside the canopy, with an angle increment of 20\% respect to the wind direction. The same percentage increase is found for the region above the canopy. The data from the weaker stratification ($Ri_{\delta}^{app}  = -0.5$, not shown in the figure) lead to a remarkably similar result for the plume direction above the canopy, while the value within the urban model is close to the neutral reference case.

\begin{figure}
	\centering
	\includegraphics[width=0.8\linewidth]{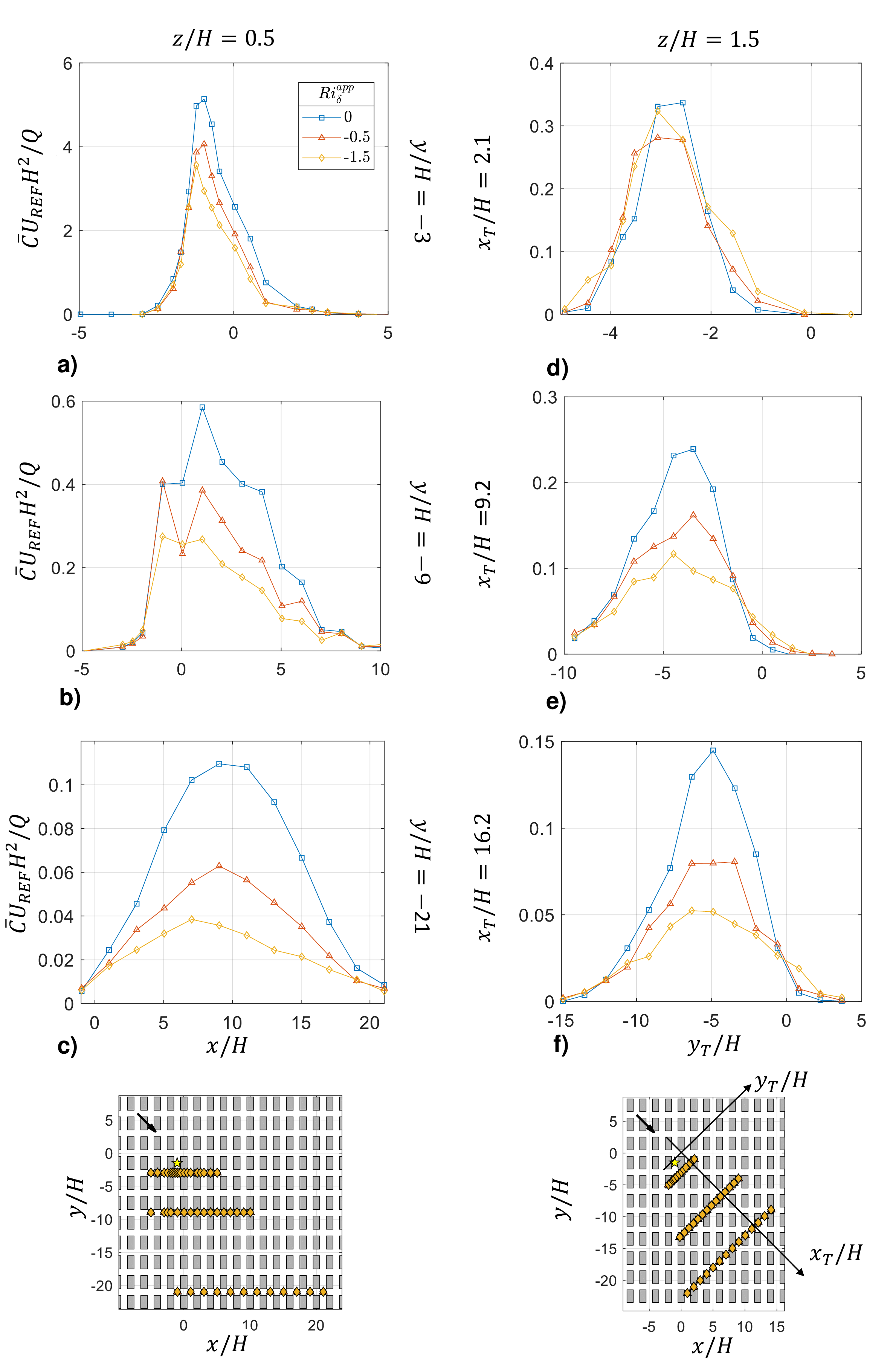}
	\caption{Lateral profiles of mean concentration inside and above the canopy for three levels of instability. The star on the maps at bottom corresponds to the source, while the other marks show the locations of the measurement points along the lateral profiles.}
	\label{fig:LateralPlumeCBL}
\end{figure}

\begin{figure}
	\centering
	\includegraphics[width=1\linewidth]{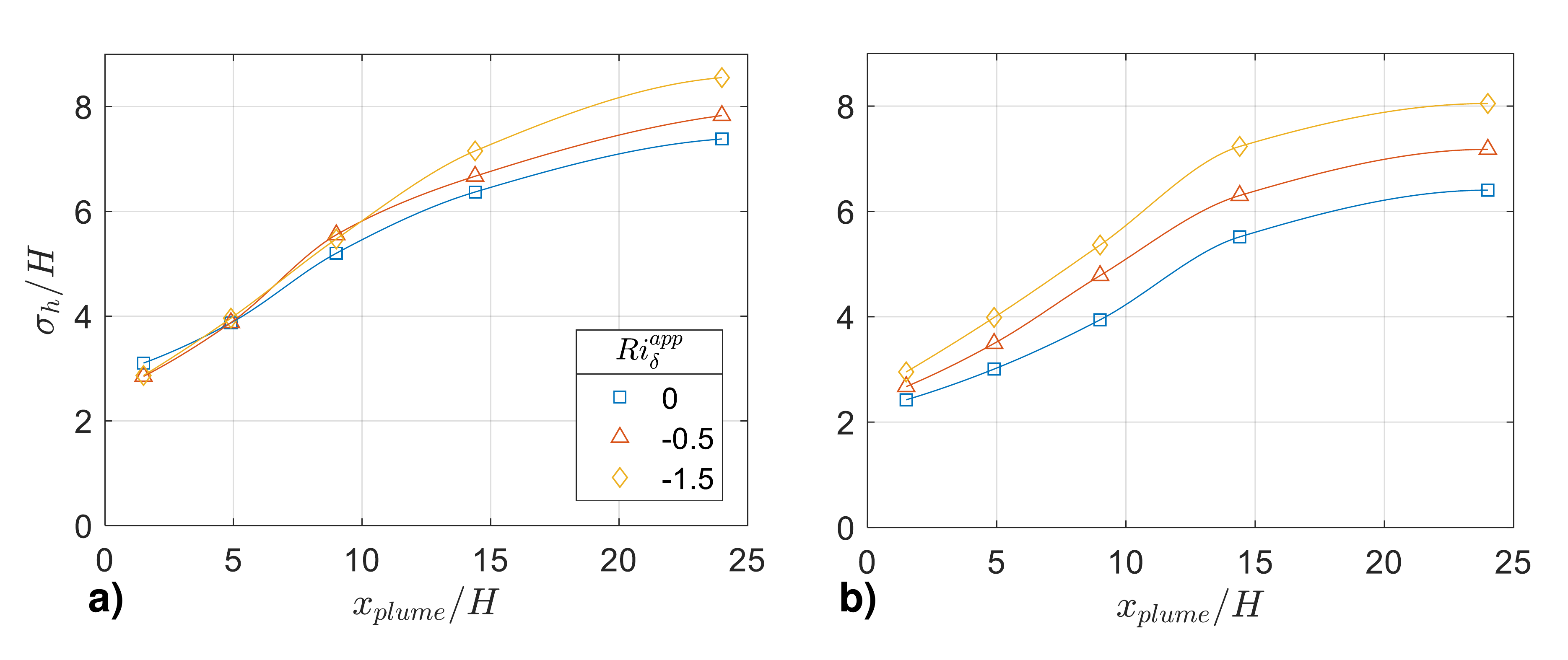}
	\caption{$\sigma_h$ for CBL and NBL varying the distance from the source at $z/H$ of 0.5 (a) and 1.5 (b).}
	\label{fig:SigmahCBL}
\end{figure}

When comparing the mean concentration values the unstable stratification effect appears opposite to what measured for the SBL. In this case, the concentration levels within the canopy are reduced almost everywhere (up to three times), as a consequence of the increased vertical exchange. This fact is better appreciable in Fig.~\ref{fig:LateralPlumeCBL}, where the lateral profiles of the two cases are shown, together with a case of intermediate instability. The results for the latter lays between the NBL and the stronger instability case. Fig.~\ref{fig:SigmahCBL} displays the computed values of $\sigma_h$, representative of the plume width. The trend shows here a clearer increase inside the canopy (after 9$H$), compared to the NBL. Above the canopy a difference is discernible throughout the plume, as it was for the SBL. The results for the intermediate instability case lie again between the NBL and the strongest instability.

\begin{figure}
	\centering
	\includegraphics[width=1\linewidth]{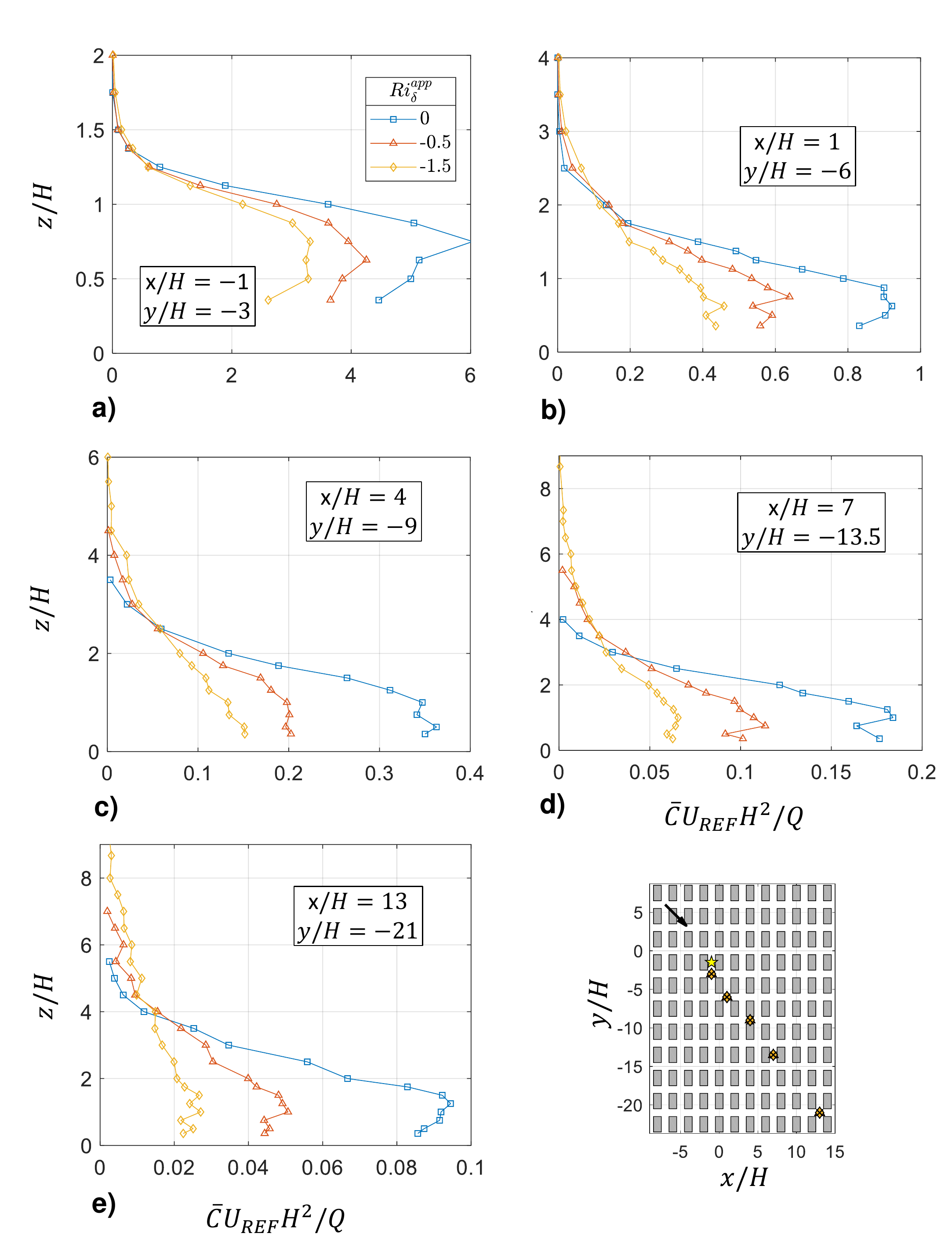}
	\caption{Vertical profiles of mean concentration approximately along the plume axis for three levels of instability. The star on the map at bottom-right corresponds to the source, while the other marks show the locations of the five vertical profiles.}
	\label{fig:VerticalPlumeCBL}
\end{figure}

\begin{figure}
	\centering
	\includegraphics[width=0.6\linewidth]{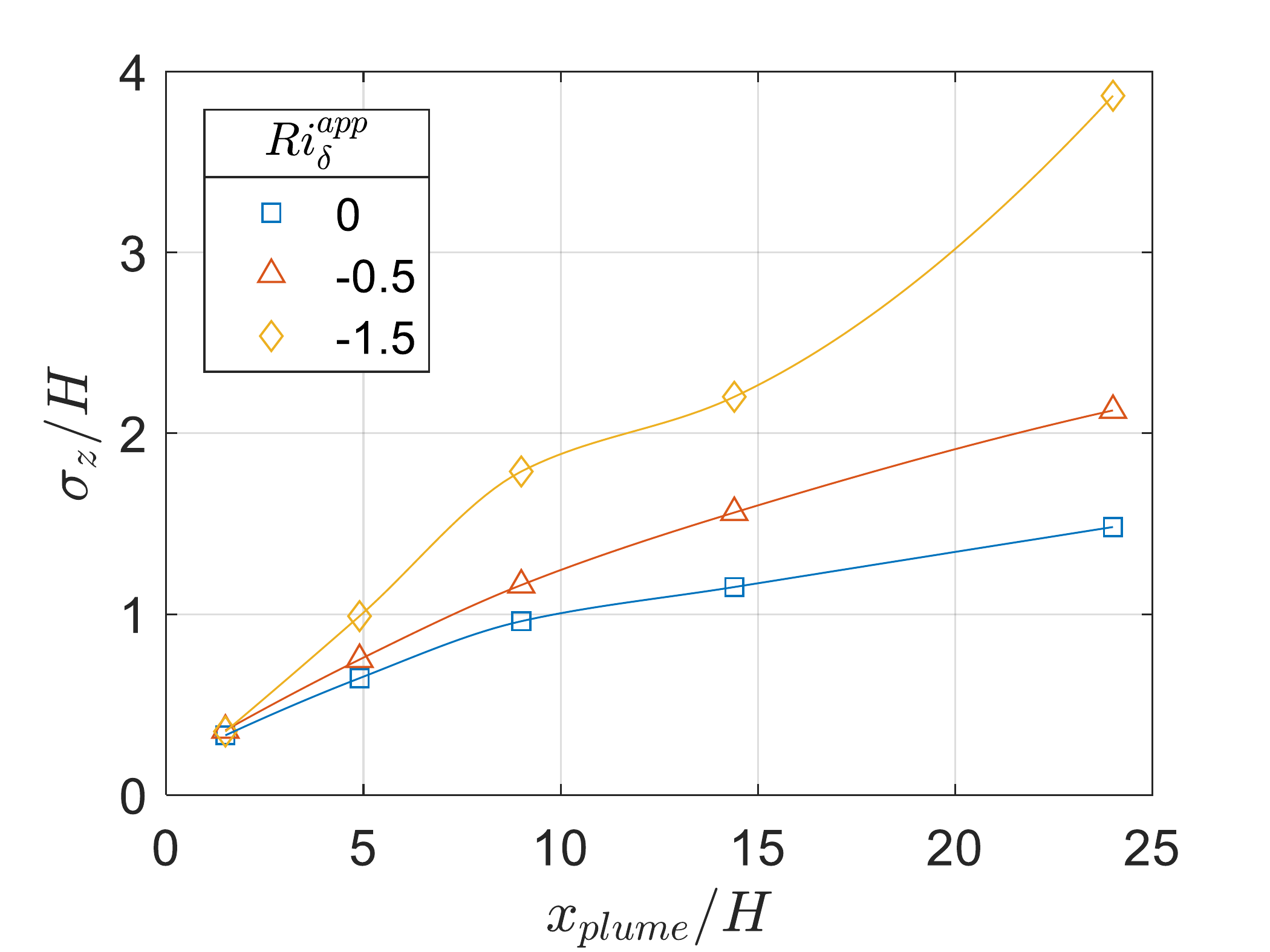}
	\caption{$\sigma_z$ for CBL and NBL varying the distance from the source.}
	\label{fig:SigmazCBL}
\end{figure}

The plume depth starts differing from $x/H = 1$, as discernible in the vertical profiles of mean concentration in Fig.~\ref{fig:VerticalPlumeCBL}. The plots clearly show lower concentrations within the canopy, compared with the neutral case (as already mentioned in the analysis of the later profiles), and higher concentrations further up. The plume, then, appears deeper, indicating that the pollutant tracer is able to penetrate deeper into the BL above the canopy, reaching a depth of more than 7$H$ at the farthest measured location, even though with very low concentration values. Such a trend is expected, since the enhanced vertical exchange due to the buoyancy forces contributes to clean the air inside the canopy, facilitating the exchange with the region above. The $\sigma_z$ plot in Fig.~\ref{fig:SigmazCBL} confirms this behaviour, with the parameter showing a clear and progressive increment after the application of unstable stratification, more evident than the variation in the plume width. Again this result is in accordance with \cite{Briggs1973} and in contrast with \cite{Kanda2016}.

\begin{figure}
	\centering
	\includegraphics[width=1\linewidth]{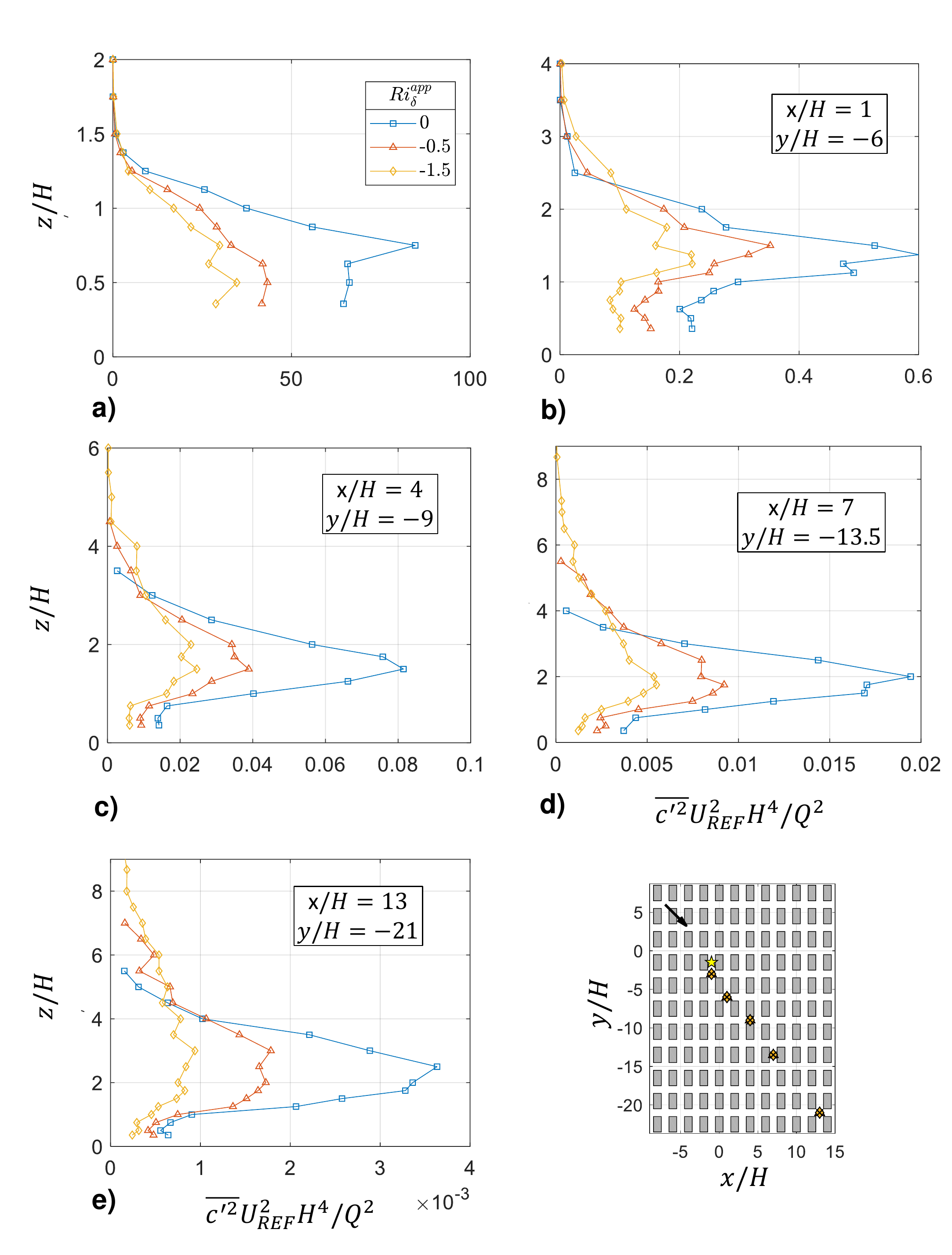}
	\caption{Vertical profiles of concentration variance approximately along the plume axis for three levels of instability. The star on the map at bottom-right corresponds to the source, while the other marks show the locations of the five vertical profiles.}
	\label{fig:VerticalPlumevarianceCBL}
\end{figure}

The concentration variance (Fig.~\ref{fig:VerticalPlumevarianceCBL}) seems to behave like described for the stable cases, varying according to the mean concentration levels.

\section{Vertical pollutant fluxes}
\label{sec:flux}
\begin{figure}
	\centering
	\includegraphics[width=1\linewidth]{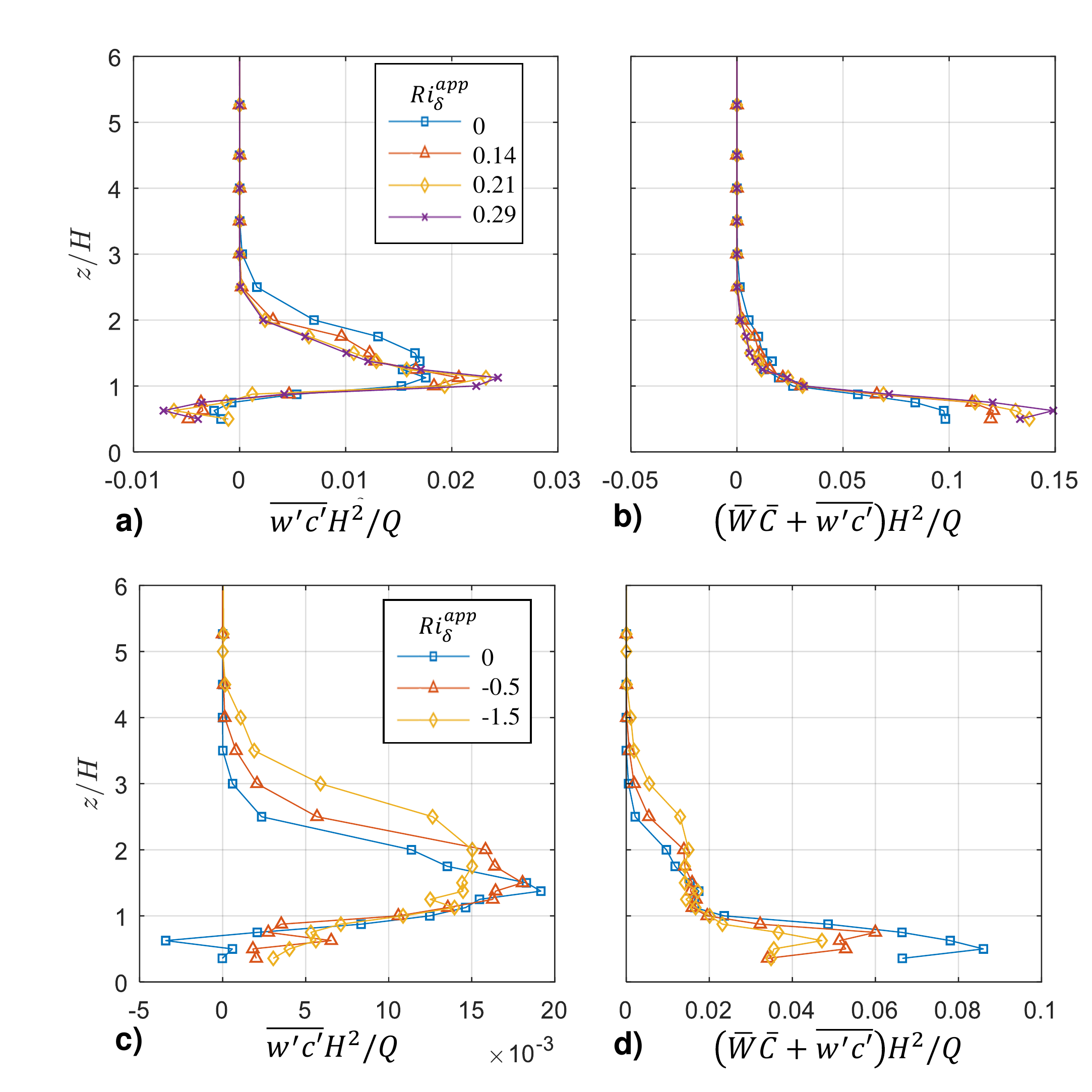}
	\caption{Vertical profiles of turbulent and total vertical pollutant flux varying the stable (a, b) and unstable stratification (c, d) at the centre of an intersection ($x/H = 1$, $y/H = -6$).}
	\label{fig:wcSBL}
\end{figure}

Fig.~\ref{fig:wcSBL} shows the graphs of vertical turbulent and total pollutant fluxes with varying stable and unstable stratification levels at a location at the centre of an intersection. For the SBL cases, inside the canopy the turbulent fluxes are close to zero (and slightly negative), while the total ones experience a peak at about 0.5$H$ (the lowest measured position), meaning that the mean pollutant fluxes are predominant there. In general, the total vertical fluxes follow the trend of the mean concentration profile, also when different levels of stratification are involved. Despite this, the turbulent fluxes experience a steep peak at roof level (or slightly above), reaching values similar to the mean fluxes. This is an important aspect because the roof level is critical in the exchange between the canopy and the upper region. Moreover, the total pollutant flux at roof level is not seen to be affected by the stratification, at least at the centre of the intersection. The fact that the total fluxes inside the canopy are larger in the stably-stratified cases despite the reduced vertical turbulence \citep[see][]{Marucci2019flow} is indicative of the predominance of the mean fluxes over the turbulent ones. Above the canopy, however, both the total and turbulent flux appear to be reduced by stratification.
In the CBL case the vertical velocity fluctuations are enhanced everywhere \citep{Marucci2019flow}. On the other hand, the concentration levels are reduced inside and above the canopy until a point (that in the case of Fig.~\ref{fig:VerticalPlumeCBL}b is at about $2H$) after which the concentration starts being larger than the NBL, hence making the plume deeper. In this situation, the vertical turbulent pollutant flux appears generally increased inside the canopy and above 1.5$H$. In the region immediately above the roof level, instead, a steep gradient seems to advantage the neutral case. That said, inside the canopy the turbulent flux remains irrelevant compared to the mean values except, again, at roof level and above, where they have the same order of magnitude.

\begin{figure}
	\centering
	\includegraphics[width=0.8\linewidth]{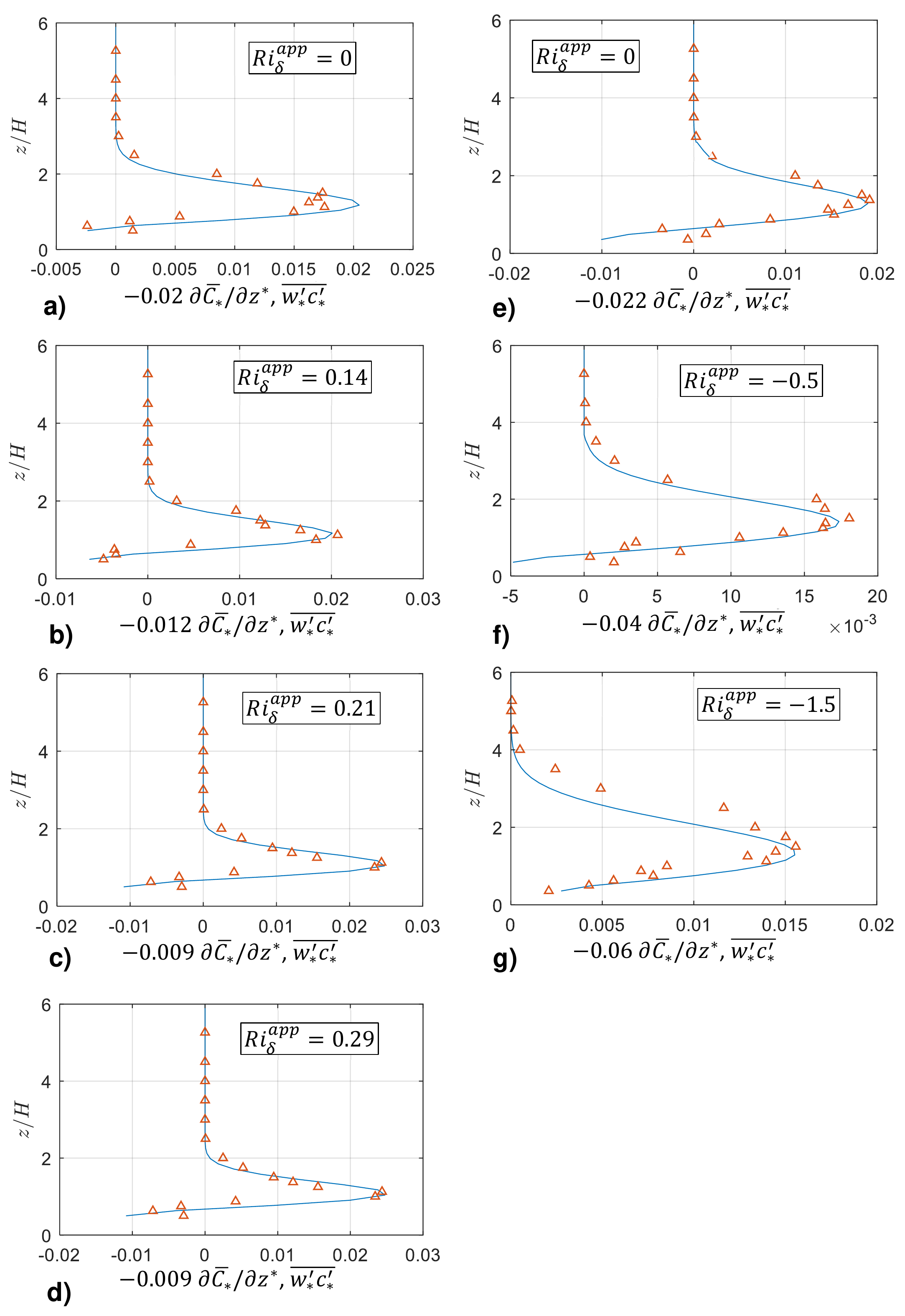}
	\caption{Vertical profiles of vertical turbulent pollutant fluxes ($x/H = 1$, $y/H = -6$) with varying stratification. The blue line is the gradient of dimensionless concentration over z/H obtained by a Gaussian fit of the mean concentration vertical profile.}
	\label{fig:wcGrad}
\end{figure}

An interesting point to analyse is the similitude between vertical turbulent pollutant flux and concentration gradient

\begin{equation}
K_z \frac{\partial \overline{C}}{\partial z} = -\overline{w'c'}
\end{equation}

\noindent where $K_z$ is a constant of proportionality (called ``eddy diffusivity'').
Such behaviour was demonstrated by \cite{Dezso-Weidinger2003}, confirmed by \cite{Carpentieri2012} for neutral stratification and it is normally used in models to compute vertical turbulent pollutant fluxes (as e.g. SIRANE, see \citealp{Soulhac2011}). Nevertheless, its validity in the SBL and CBL cases was still questioned. In Fig.~\ref{fig:wcGrad} profiles of vertical turbulent pollutant fluxes are plotted and compared with the concentration gradient profiles obtained from a Gaussian fit of the mean concentration. The proportionality in this case is evident, though the constant of proportionality seems to vary. In particular, it tends to increase with unstable stratification and decrease with stable, ranging from 0.009 to 0.06. A variability depending on the location and mechanical turbulence was found by \cite{Carpentieri2012} and it is confirmed here (the constant reaching a value of 0.14 in case of stronger stratification, see Tab.~\ref{table:Kz}). Of course, the analysis in this case is based on very specific locations at the centre of the intersection or the street canyons. The numerical simulation results by \citet{Fuka2018} on the neutral case show that the eddy diffusivity can even be negative at certain locations.

\begin{table}
	\caption{Values of $K_z$ varying stratification and location}
	\centering
	\label{table:Kz}
	\scalebox{0.75}{
		\begin{tabular}{l c c | c c c c}
			\toprule
			\multicolumn{3}{c|}{\textbf{Stability Case}} & \multicolumn{4}{c}{$K_z$} \\
			&  &  & x/H=1 & x/H=4 & x/H=7 &  \\
			$\mathrm{Ri_\delta^{app}}$ & $\mathrm{Ri_\delta}$ & $\mathrm{\delta/L}$ & y/H=$-$6 & y/H=$-$9 & y/H=$-$13.5 & Mean \\
			\midrule
			0 (SBL) & 0 & 0 & 0.020 & 0.030 & 0.035 & 0.028 \\
			0.14 & 0.12 & 0.40 & 0.012 & 0.018 & 0.018 & 0.016 \\
			0.21 & 0.19 & 0.62 & 0.010 & 0.014 & 0.015 & 0.013 \\
			0.29 & 0.25 & 0.69 & 0.009 & 0.012 & 0.012 & 0.011 \\
			\midrule
			0 (CBL) & 0 & 0 & 0.022 & 0.030 & 0.035 & 0.029 \\
			$-$0.50 & $-$0.35 & $-$0.51 & 0.040 & 0.060 & 0.080 & 0.060 \\
			$-$1.50 & $-$0.91 & $-$1.09 & 0.060 & 0.100 & 0.140 & 0.100 \\
			\bottomrule
	\end{tabular}}
\end{table}

\begin{figure}
	\centering
	\includegraphics[width=\linewidth]{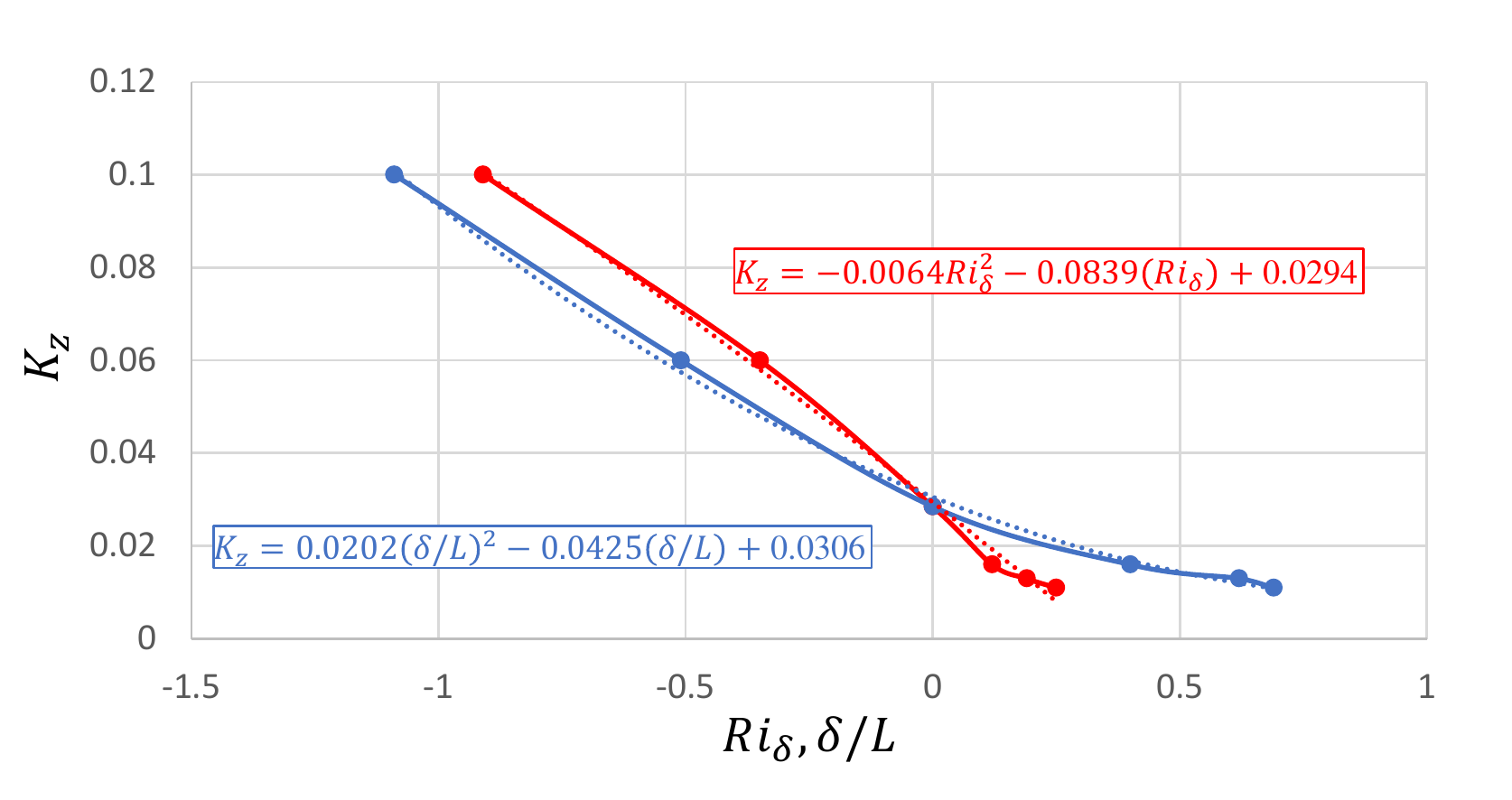}
	\caption{Mean value of $K_z$ at three locations plotted against $Ri_\delta$ or $\delta/L$. Dotted lines are obtained by fitting the experimental data with a polynomial curve.}
	\label{fig:KzVSL}
\end{figure}

In Fig.~\ref{fig:KzVSL} the values of the mean $K_z$ from Tab.~\ref{table:Kz} are plotted against $Ri_\delta$ and $\delta/L$. A parametrisation is attempted by means of a polynomial fitting of the second order (also shown in the figure)

\begin{equation}
K_z\left(\delta/L\right) = 0.0202\left(\delta/L\right)^2 - 0.0425\left(\delta/L\right) + 0.0306
\end{equation}
\begin{equation}
K_z\left(Ri_\delta\right) = -0.0064 Ri_\delta^2 - 0.0839 Ri_\delta + 0.0294
\end{equation}

\section{Conclusion}
\label{sec:Conclusion}

Wind tunnel experiments were conducted to study the impact of atmospheric stratification on flow and dispersion within and over a regular array of rectangular buildings at a 45$^\circ$ wind angle. Three stable and two convective incoming boundary layers were tested with a Richardson number ranging from $-$1.5 to 0.29. Dispersion measurements were carried out using propane released from a point source within the urban model as tracer gas, sampled using a fast FID probe. Simultaneous velocity and temperature measurements were also taken \citep{Marucci2019flow}. The dispersion plume was sampled in and above the canopy by means of lateral and vertical profiles.

The results of the pollutant dispersion measurements show that the stratification (either stable or unstable) effect on the plume width is significantly lower than the effect on the vertical profiles (as also indicated by \citealp{Briggs1973}, but in contrast with the results by \citealp{Kanda2016}). Stable stratification did not affect the plume central axis inside the canopy, but in the unstable case the axis appeared to deviate from the neutral case direction. Above the canopy both stratification types caused an increase in the plume deflection angle compared to the neutral case. Measured concentrations in stable stratification were up to two times larger in the canopy compared to the neutral case, the opposite for the convective stratification (which are up to three times lower). Vertical turbulent pollutant fluxes have been found to be only slightly affected by stratification, but without significant changes in the general trend. Mean pollutant fluxes in the canopy remain predominant close to the source, even though at roof level and above turbulent and mean fluxes have the same order of magnitude. The proportionality between the vertical turbulent fluxes and the vertical mean concentration gradient (base of the K-theory) is confirmed also in the stratified cases.

The experimental data produced during this work, to the authors' knowledge, are the most comprehensive available so far for urban flow and dispersion studies in presence of atmospheric stratification and they may help developing new mathematical models and parametrisation, as well as validating existing and future numerical simulations.

The tested boundary layer stratification levels ranged from weakly stable to weakly unstable. Despite the fact that more extreme conditions may create more dramatic effects on the aerodynamic and dispersion properties, it should be noted that in urban areas extreme stratifications are normally quite uncommon (excluding locations at larger latitudes were very stable conditions may occur even in rural or urban areas). \citet{Wood2010} showed, for example, that in London during a long experimental campaign, the most frequent cases are the ones characterised by lower stratification level, with the region in the range $-1<z'/L<1$ occurring for about 75\% of the times, both during night and day (where the reference height $z'$ represents the difference between the measurement height, 190.6~m, and the displacement height over the city). Unfortunately, the boundary layer depth for each of these cases was not indicated \citet{Wood2010}, so a comparison with the wind tunnel data is hard, but considering a typical scaling ratio of 1/200 the resulting Monin-Obukhov length values at full scale for the experimental data in the present work are of the order of $\pm200$~m (hence approximately in the range of $-1<z'/L<1$ compared to the London data, and so covering 75\% of the actual cases).

Future experiments might include different wind directions and different urban geometries. Given the significant impact of stratification on the vertical spread of the pollutant plume, it would be particularly interesting to apply the methodology developed in this paper to urban geometries that include very tall buildings \citep{Fuka2018,Hertwig2019,Aristodemou2018}.

\section*{Acknowledgments}
The authors are grateful for the financial support by the EPSRC (grant EP/P000029/1) and by the Department of Mechanical Engineering Sciences (University of Surrey).

\section*{Data availability}
Wind tunnel data are available at \href{https://doi.org/10.6084/m9.figshare.8320007}{https://doi.org/10.6084/m9.figshare.8320007}.






\bibliography{StratEnFlo_array_dispersion}


\end{document}